\documentclass[sigconf]{acmart}
\usepackage{tikz}
\usetikzlibrary{arrows.meta,positioning,fit}
\usepackage{subcaption}
\usepackage{graphicx}
\usepackage{caption}
\usepackage{color,colortbl}
\usepackage{colortbl}

\usepackage{amsmath}

\usepackage{xcolor}
\usepackage{soul}
\usepackage{siunitx}
\usepackage{booktabs}
\usepackage{enumitem}
\usepackage{multirow}
\usepackage{pifont}
\usepackage{xspace}
\usepackage{hyperref}
\usepackage{cleveref}
\usepackage{balance}
\usepackage{listings}
\usepackage{tabularx}
\usepackage{fancyhdr} 
\usepackage{lipsum}
\usepackage{tcolorbox}
\usepackage{tabu}

\newcounter{RQACounter}

\newcommand{\RS}[2]{%
\begin{tcolorbox}[width=\columnwidth]
\vspace{-1mm}
\textbf{Result {#1}:~}{\emph {#2}}%
\vspace{-1mm}
\end{tcolorbox}
}

\AtBeginDocument{%
  }

\setcopyright{acmlicensed}
\copyrightyear{2018}
\acmYear{2018}
\acmDOI{XXXXXXX.XXXXXXX}
\acmConference[Conference acronym 'XX]{Make sure to enter the correct
  conference title from your rights confirmation email}{June 03--05,
  2018}{Woodstock, NY}
\acmISBN{978-1-4503-XXXX-X/2018/06}

\title{The Layout \textit{Is} the Model: On Action-Item Coupling in Generative Recommendation
}

\author{Xiaokai Wei}
\email{xwei@roblox.com}
\affiliation{\institution{Roblox} \city{San Mateo, California} \country{USA}}
\author{Jiajun Wu}
\email{jiajunwu@roblox.com}
\affiliation{\institution{Roblox} \city{San Mateo, California} \country{USA}}
\author{Daiyao Yi}
\email{dyi@roblox.com}
\affiliation{\institution{Roblox} \city{San Mateo, California} \country{USA}}
\author{Reza Shirkavand}\authornote{Contribution was made while being a graduate collaborator at Roblox.}
\email{rezashkv@umd.edu}
\affiliation{\institution{University of Maryland} \city{College Park, Maryland} \country{USA}}
\author{Michelle Gong}
\email{mgong@roblox.com}
\affiliation{\institution{Roblox} \city{San Mateo, California} \country{USA}}

\renewcommand\shortauthors{Wei et al.} 

\begin{abstract}

Generative Recommendation (GR) models treat a user’s interaction history as a sequence to be autoregressively predicted. When both items and actions (e.g., watch time, purchase, comment) are modeled, the \emph{layout}---the ordering and visibility of item/action tokens---critically determines what information the model can use and how it generalizes. We present a unified study of token layouts for GR grounded in first principles: (P1) maximize item/action signal in both input/output space, (P2) preserve the conditioning relationship ``action given item'' and (P3) no information leakage. 

While \emph{interleaved} layout (where item and action occupy separate tokens) naturally satisfies these principles, it also bloats sequence length with larger training/inference cost. On the \emph{non-interleaved} front, we design a novel and effective approach, \emph{Lagged Action Conditioning} (LAC), which appears strange on the surface but aligns well with the design principles to yield strong accuracy. Comprehensive experiments on public datasets and large-scale production logs evaluate different layout options and empirically verifies the design principles. Our proposed non-interleaved method, \emph{LAC}, achieves competitive or superior quality at substantially lower FLOPs than interleaving. Our findings offer actionable guidance for assembling GR systems that are both accurate and efficient.

\end{abstract}



\received{20 February 2007}
\received[revised]{12 March 2009}
\received[accepted]{5 June 2009}

\begin{document}
\maketitle

\section{Introduction}

Generative Recommendation (GR) \citep{Rajput2023RecommenderSW,Zhai2024ActionsSL, Zhou2025OneRecTR, Khrylchenko2025ScalingRT} treats a user’s interaction history as a sequence to be autoregressively modeled: past tokens condition a decoder that predicts future tokens, which are then mapped to recommended items (and optionally their actions). By operating over a compact token vocabulary decoupled from catalog size, GR offers appealing scalability and memory efficiency while leveraging sequence modeling progress from language models \citep{Kang2018SasRec,Zhou2017DeepIN}. Yet \emph{how} we arrange information in that sequence, i.e., the \textbf{layout} of item and action tokens and their allowed attention, critically determines what the model can learn, what it can safely condition on, and at what computational cost.

In traditional Deep Learning Recommendation System (DLRM) \cite{Cheng2016WideD, Wang2020DCNVI}, the training example construction is relatively straightforward. Each example is a fixed \emph{(features, label)} pair. Everything in the feature vector is, by construction, legal to use for predicting the label; leakage is easy to avoid (do not include the target or a deterministic proxy). Compute is predictable (embedding lookups + MLP), and the interface is simple: build features $\rightarrow$ train classifier/regressor. 

In constrast, generative recommendation has stronger scaling laws \cite{Zhai2024ActionsSL}\cite{Han2025MTGR}\cite{Zhou2025OneRecTR} but layout-sensitive. GR reframes supervision as a token sequence: (1) what used to be ``features'' becomes the prefix (observed history); (2) what used to be the ``label'' becomes the next token under a causal mask. That forces deliberate choices about (i) which symbols (items/actions) become tokens, (ii) their order, and (iii) which earlier tokens each position may attend to (the visibility policy). These choices, the \emph{layout}, control both what the model can learn and the efficiency of training/inference.

Hence, in generative recommendation (GR), the \emph{layout is a hidden degree of freedom}—and we argue layout should be treated as a \emph{first-class design axis}. While in DLRM legality and leakage avoidance are baked into the feature vector, in GR they are created by the layout of tokens. The same backbone and data can vary in quality/latency purely from layout choices, so we must design to clear principles, not stumble into them.

From first principles, an effective layout should: \textbf{(i)} maximize useful information in the conditioning set without leakage, \textbf{(ii)} preserve the natural causal direction ``action given item'', while keeping sequence length short for efficiency. A growing line of work \citep{Zhai2024ActionsSL,Han2025MTGR,Zhang2025KillingTB} interleaves item and action tokens, e.g., \(\texttt{I}_1,\texttt{A}_1,\texttt{I}_2,\texttt{A}_2,\ldots\), so that the action at step \(t\) can attend to its same-step item \(\texttt{I}_t\). This design captures (ii) causal direction ``\emph{action given item}'' but it doubles sequence length and increases the $O(T^2)$ attention cost. Recent variants (e.g., dual-flow training \cite{Guo2025ActionIA}) mitigate some overhead by using two mirrored flows during training and a single flow at inference, but they still complicate implementation and deployment. At the other extreme, dropping actions preserves efficiency but violates (i) by discarding strong predictors such as recent engagement intensity.

In this paper, we focus on efficient non-interleaved layouts and aim to close the gap by systematically investigating the implications of different layouts. The result is effective \emph{non-interleaved} designs that are leakage-safe, causally faithful, and compute-efficient, preserving the crucial action-given-item dependency without paying the full interleaving penalty. This provides a principled recipe for assembling GR systems that meet industrial constraints while retaining the expressivity needed for high-quality ranking.

Our contributions can be summarized as follows:
\begin{itemize}
\item \textbf{Practical pricinples for layout-centric GR design} We recognize the importance of token choreography in GR and analyze the tension of action-item coupling in GR model layout. We formalize several principles to guide layout design, especially for non-interleaved setting. 
\item \textbf{Novel and effective layout.} We introduce a simple and effective layout that preserve action-given-item conditioning, while keeping sequence length \(T\) and avoiding specialized implementations. Though the proposed \emph{Lagged-Action-Conditioning} (LAC) approach seems to have a mismatched input, we show how the emergent specialized attention pattern makes this design a sensible and elegant one. 
\item \textbf{Evaluation with comprehensive ablations.} We perform extensive evaluation on both public and proprietary industrial datasets to demonstrate the superior performance of our approach. We also instantiate $>10$ action/item layout variants to empirically validate the design principles we developed. We also study whether the competitive performance of LAC persists when scaling up model sizes and with more than one tasks.
\end{itemize}

\section{Related Work}

Deep learning based recommendation system (DLRM) \cite{Cheng2016WideD, Wang2020DCNVI} typically utilize feature crossing network (e.g., DCNv2 \cite{Wang2020DCNVI} to fuse multiple input features. Recently, generative recommendation \citep{Rajput2023RecommenderSW, Han2025MTGR, Zhou2025OneRecTR, Zhang2025KillingTB, Huang2025TowardsLG, Khrylchenko2025ScalingRT, Chen2025PinFM, Shirkavand2025CatalogNativeLS} have emerged as a promising paradigm. The generative recommenders predominatly employs a decoder-only \cite{radford2019language} autoregressive model. Similar to the pretraining objectives of LLM, typically next-item-prediction (or multi-token-prediction) are employed. To better generalize to less popular items, certain work \cite{Rajput2023RecommenderSW, Zhou2025OneRecTR, Zhu2024CoSTCQ, Wang2024LearnableIT, Li2024SemanticCH} also explored constructing semantic ID for items based on the item content (e.g., textual/visual or other modalities) instead of (hashed) item IDs. 

To enable the capability of scoring items, interleaved paradigm is adopted by multiple generative recommendation models such as \cite{Zhai2024ActionsSL} \cite{Han2025MTGR} \cite{Zhang2025KillingTB}. The downside is also obvious: input sequence is inflated to twice the original length, which could incur higher FLOPs for training and inference. Considering such drawback of interleaved approaches, recent models start to pay attention to non-interleaved approach. For example, \cite{Huang2025TowardsLG}, \cite{Khrylchenko2025ScalingRT} and \cite{Chen2025PinFM} try to employe non-interleaved layout for unbloated sequence length. However, to our best knowledge, no prior work has systematically discussed the trade-offs of different layout designs. As we will discuss in the following sections, more optimal design trade-off exists, which makes non-interleaved approach have comparable accuracy as interleaved approaches.

\section{Preliminaries}
\subsection{Formulation}
We model each user $u$ as a sequence of timestamped interactions $\{e_t\}_{t=1}^{T_u}$ with $e_t=(i_t,a_t)$, where $i_t\in\mathcal{I}$ is an item (e.g., a video/product/game) and $a_t\in\mathcal{A}$ is an action/engagement signal (e.g., watch time, like, comment, purchase value or review rating). A \emph{layout} $\mathcal{L}$ maps interactions to a token sequence
\[
\mathbf{y}^{(u)}_{1:L_u} \;=\; \mathcal{L}_{\text{tokens}}\big(\{(i_t,a_t)\}_{t=1}^{T_u}\big),
\]
together with a causal visibility mask $\mathcal{M}^{(u)}=\mathcal{L}_{\text{mask}}$.
A GR model maximizes the AR likelihood under the mask:
\[
\max_{\theta}\ \sum_{u}\ \sum_{k=1}^{L_u} \log p_{\theta}\!\big(y_k^{(u)} \,\big|\, y_{<k}^{(u)};\,\mathcal{M}^{(u)}\big),
\]
where $y_k$ are \emph{layout-defined} tokens and the mask enforces causal validity (no leakage).

\paragraph{Retrieval and scoring in the AR view.}
At time $t$ (with prefix $y_{\le k}$ corresponding to history up to $t$), \emph{retrieval} ranks candidates $c\in\mathcal{I}$ by
\[
S_{\text{ret}}(c \mid y_{\le k}) \;=\; p_{\theta}\!\big(i_{t+1}{=}c \,\big|\, y_{\le k}\big),
\]
while \emph{ranking/scoring} is made \emph{target-aware} by additionally modeling
\[
S_{\text{rank}}(c \mid y_{\le k}) \;\propto\; \mathbb{E}_{\theta}\!\big[a_{t+1}\,\big|\, i_{t+1}{=}c,\, y_{\le k}\big]
\quad\text{or}\quad
p_{\theta}\!\big(a_{t+1}\,\big|\, i_{t+1}{=}c,\, y_{\le k}\big),
\]
depending on whether $a_{t+1}$ is continuous (e.g., watch time and purchase value) or discrete (e.g., like/comment/share).

\subsection{Retrieval in generative recommendation.}
The \emph{retrieval task} constructs a small candidate set $\mathcal{C}_{u,t}\subset\mathcal{I}$ by ranking items with the next-item distribution $p_\theta(i_{t+1}\mid y_{\le k})$ induced by the chosen layout and mask. Training commonly uses the next-item objective with sampled softmax or contrastive NCE, tying output weights to the item embedding table for stability \cite{radford2019language}. At inference, the context state $h_t$ is computed once and items are scored via a dot product $h_t W_I^\top$ (or a calibrated variant), selecting top-$K$ as candidates. While \emph{item-only} GR can perform retrieval task, layouts that inject actions often improve $p(i_{t+1}\mid\cdot)$ by incorporating action value as additional context.


\subsection{Scoring/Ranking in generative recommendation.}
The \emph{scoring task} (or referred to as \emph{ranking} in some existing work \cite{Zhai2024ActionsSL}) estimates a user’s response to a specific item given context—e.g., $p(a_{t+1}\mid i_{t+1}, \text{history})$ or $\mathbb{E}[a_{t+1}\mid i_{t+1}, \text{history}]$, where $a$ can be watch time, like, purchase value, playtime, etc. Scoring operates after retrieval (in industrial setting the retrieval stage typically employs multiple sources/approaches to get diverse candidates) has produced $\mathcal{C}_{u,t}$. The model assigns a \emph{target-aware} score to each $c\in\mathcal{C}_{u,t}$ so the system can rank items by fusion of multiple business objectives. Unlike next-item likelihoods, scoring captures value-conditioned-on-item and is central across different domains:
\begin{itemize}
  \item \textbf{E-commerce:} rank by purchase probability or expected GMV ($\text{price}\times p(\text{purchase})$). 
  \item \textbf{Short-video:} rank by expected watch time/completion ratio and other types of engagement (e.g., comment/like/share) for retention and session health.
  \item \textbf{Games:} prioritize by expected session duration or in-game purchase.
  \item \textbf{Feeds:} combine calibrated predictions on multiple objectives (e.g., click/dwell time/thumbs-up/quality) under diversity/fatigue/policy constraints.
\end{itemize}
Layouts that enable \emph{action-given-item} (e.g., non-interleaved patched conditioning) support fast per-candidate evaluation by reusing the trunk state. This is crucial in modern recommendation systems as it opens room for inference optimization. It is also one important reason why we aim to investigate layout candidates that might lead to stronger scoring/ranking performance.

\subsection{Item-only next-item training (retrieval only).}
Choose a layout with \emph{item tokens only}:
\[
\mathbf{y}^{(u)}_{1:L_u} \;=\; (i^{(u)}_1,\,i^{(u)}_2,\,\ldots,\,i^{(u)}_{T_u}) \quad\Rightarrow\quad L_u=T_u.
\]
Train with the next-item objective
\[
\mathcal{L}_{\text{item}} \;=\; -\sum_{u}\sum_{t=1}^{T_u-1} \log p_{\theta}\!\big(i^{(u)}_{t+1}\,\big|\, i^{(u)}_{\le t}\big).
\]
This supports \emph{retrieval} via $S_{\text{ret}}$ but does not yield a target-aware $S_{\text{rank}}$, since actions $\{a_t\}$ are neither tokens nor conditionals in $\mathbf{y}$.

\subsection{Interleaved item-action: joint prediction for retrieval \emph{and} ranking.}
Use an interleaved layout with two tokens per interaction
\[
\mathbf{y}^{(u)}_{1:L_u} \;=\; \big(i^{(u)}_1, a^{(u)}_1,\, i^{(u)}_2, a^{(u)}_2,\,\ldots,\, i^{(u)}_{T_u}, a^{(u)}_{T_u}\big),
\quad L_u \approx 2T_u,
\]
and a mask that (i) is causal over time and (ii) lets the action token at step $t{+}1$ attend to the \emph{same-step} item token $i_{t+1}$ (teacher forcing). The AR loss decomposes as
\[
\mathcal{L}_{\text{IL}} \;=\; -\sum_{u}\sum_{t=1}^{T_u-1}
\Big[
\underbrace{\log p_{\theta}\!\big(i^{(u)}_{t+1}\,\big|\, y^{(u)}_{\le k(t)}\big)}_{\text{retrieval}}
\;+\;
\underbrace{\log p_{\theta}\!\big(a^{(u)}_{t+1}\,\big|\, y^{(u)}_{\le k(t)},\, \color{blue}{i^{(u)}_{t+1}}\big)}_{\text{target-aware ranking}}
\Big],
\]
where $k(t)$ indexes the position of $a_t$ in $\mathbf{y}$. This enables both tasks but doubles $L$ (higher $O(L^2)$ cost). 

Keep one token per interaction ($L_u=T_u$) but preserve target-aware conditioning via the \emph{layout} and heads. Keeping $L{=}T$ improves throughput and permits longer contexts under fixed compute, a practical edge that translates into accuracy gains.

 \begin{figure*}[h]
 \centering
  \includegraphics[width=0.9\textwidth, trim = {0mm 16mm 0mm 0mm}, clip]{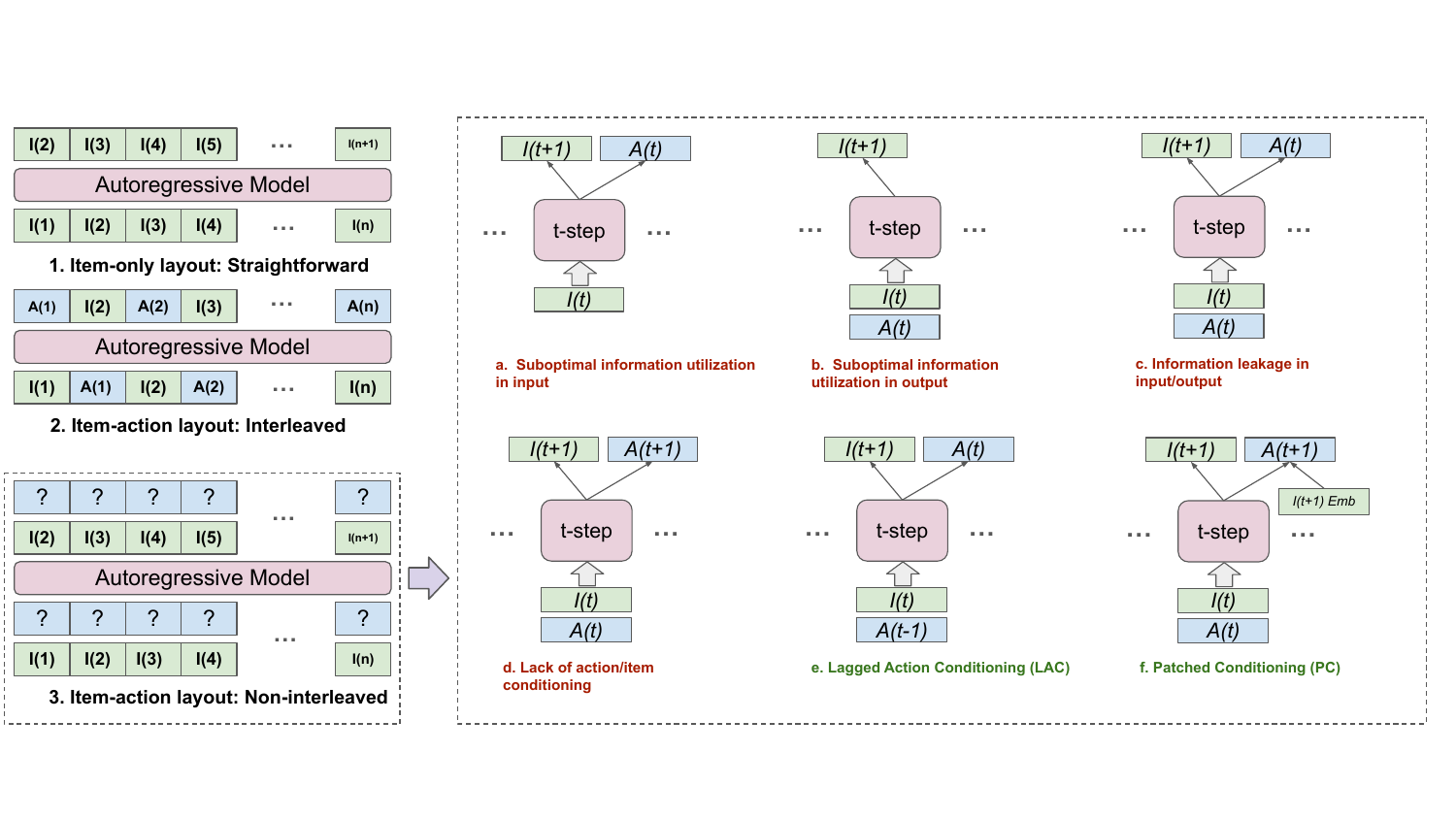}
 \caption{Item-action layout in GR. a-d are examples that violate the design pricinples}
 \label{figure:layout_option}
 \end{figure*}

\section{Methodology: Layout Design for GR}

\subsection{First principles for layout design}
In this subsection, we discuss the principles that could guide the layout design of GR models, as well as concrete examples that respect or violate these principles. Figure \ref{figure:layout_option} (a)-(d) summarizes the common anti-patterns for layout design.

\textbf{P1. Maximize useful information in the input/output space}
 If leakage is avoided, more predictive context per step is strictly better. For example, at step $t$, conditioning on $(i_t,a_t)$ to predict $i_{t+1}$ dominates conditioning on $i_t$ alone, because $a_t$ summarizes the user’s engagement with $i_t$.
  \begin{itemize}
    \item \textbf{Weak.} 
      \emph{Input: $i_t$ $\rightarrow$ Predict: $i_{t+1}$} \; (legal and efficient, but discards $a_t$).
    \item \textbf{Better but not full utilization.} 
      \emph{Input: $i_t{+}a_t$ $\rightarrow$ Predict: $i_{t+1}$} \; (uses engagement with $i_t$ to inform the next-item; more informative than item-only but fails to utilize action as auxiliary supervision in output space).\\
      \emph{Input: $i_t$ $\rightarrow$ Predict: $i_{t+1}{+}a_t$} \; (uses both item and action as supervision but only feeding item $i_t$ without engagement intensity is suboptimal).
    \item \textbf{Caveat (P1 vs P2).}
      \emph{Input: $i_t{+}a_t$ $\rightarrow$ Predict: $(i_{t+1},a_{t+1})$} \; (rich context for both targets, but see P2: $a_{t+1}$ cannot condition on $i_{t+1}$ in this organization).
  \end{itemize}

\textbf{P2. Preserve the natural causal direction ``action given item''}
Many business metrics depend on $p(a_{t+1}\mid i_{t+1}, \text{history})$. A good layout allows the model to condition $a_{t+1}$ on $i_{t+1}$ (not the other way around) during training  and inference.
  \begin{itemize}
    \item \textbf{Satisfy (same-step variant).}
      \emph{Input: $i_t$ $\rightarrow$ Predict: $a_t$} \; (models ``action given item'' at time $t$; useful when the business objective is same-step).
    \item \textbf{Violate (no target-aware conditioning).}
      \emph{Input: $i_t$ $\rightarrow$ Predict: $a_{t+1}$} \; (action is predicted without access to $i_{t+1}$).\\
      \emph{Input: $i_t{+}a_t$ $\rightarrow$ Predict: $a_{t+1}$ } \; (still lacks $i_{t+1}$ in the conditioning set).\\
      \emph{Input: $i_t$ $\rightarrow$ Predict: $(i_{t+1},a_{t+1})$} \; (both predicted from the same hidden state; $a_{t+1}$ cannot see $i_{t+1}$).
  \end{itemize}

\textbf{P3. Be leakage-safe.}
  Do not ask the model to predict a variable that is already present (explicitly or deterministically) in the visible prefix.
  \begin{itemize}
    \item \textbf{Satisfy.}
      \emph{Input: $i_t$ $\rightarrow$ Predict: $a_t$} \; (no $a_t$ in input).\\
      \emph{Input: $i_t$ $\rightarrow$ Predict: $i_{t+1}$} \; (future item only).\\
      \emph{Input: $i_t{+}a_t$ $\rightarrow$ Predict: $i_{t+1}$} \; ($a_t$ pertains to time $t$, target is $t{+}1$).
    \item \textbf{Violate.}
      \emph{Input: $i_t{+}a_t$ $\rightarrow$ Predict: $a_t$} \; (direct self-label leakage).\\
      \emph{Input: $i_t{+}a_t$ $\rightarrow$ Predict: $(a_t, i_{t+1})$} \; (leak on $a_t$; $i_{t+1}$ is fine).\\
      \emph{Input: $i_t{+}a_t$ $\rightarrow$ Predict: $(a_t, i_{t+1}, a_{t+1})$} \; (leak on $a_t$; additionally, $a_{t+1}$ lacks access to $i_{t+1}$, violating P2).
    
  \end{itemize}

\subsection{Patched Conditioning}
Recognizing the lack of conditioning on $i_{t+1}$ when using a single hidden state to predict both $i_{t+1}$ and $a_{t+1}$ from inputs like $(i_t, a_t)\!\rightarrow\!(i_{t+1}, a_{t+1})$, one might add conditioning on the item as a \emph{patch}. This can be done as late fusion between the target item embedding and the autoregressive model’s last-layer hidden state. \cite{Khrylchenko2025ScalingRT} employs this way of conditioning and we refer to such condioning style as \emph{Patched Conditioning} (PC). Concretely, one can factor the joint next-step objective as

\[
p\!\left(i_{t+1}, a_{t+1} \mid \mathrm{hist}_t\right)
= p\!\left(i_{t+1} \mid \mathrm{hist}_t\right)\,
  p\!\left(a_{t+1} \mid i_{t+1}, \mathrm{hist}_t\right).
\]

and train with
\[
\mathcal{L}
= -\sum_t \Big[
  \log p\!\left(i_{t+1}\mid \mathbf{h}_t\right)
  + \log p\!\left(a_{t+1}\mid i_{t+1}, \mathbf{h}_t\right)
\Big].
\]

Architecturally, the method separates concerns. A non-interleaved trunk produces a compact context state $h_t$ (unbloated sequence, low $O(T^2)$ cost). The item head estimates $p(i_{t+1}\mid \mathrm{hist}_t)$, while a lightweight \emph{coupler} $g(h_t,e(i))$ injects the candidate’s representation into the action head to realize $p(a_{t+1}\mid i,\mathrm{hist}_t)$. During training, teacher forcing supplies $e(i_{t+1})$ to fit the conditional; at inference, one need to plug in $e(\hat i_{t+1})$ for the decoded item or evaluate $\{g(h_t, e(c))\}_{c\in\mathcal{C}}$ in parallel for per-candidate scoring without re-running attention. Accordingly, PC enables ``action-given-item'' conditioning for inference as the action estimate is explicitly conditioned on the specific candidate item.

\subsection{Proposed: Lagged Action Conditioning}
We propose a novel non-interleaved layout that offers a better trade-off for item-action coupling in generative recommendation: 
\[
\boxed{\ (i_t,\, a_{t-1}) \;\longrightarrow\; (i_{t+1},\, a_t)\ }.
\]
where we pair the $t$-th item with lagged action, i.e., action value on previous item ($i_{t-1}$), at each step. We refer to the proposed approach as \emph{Lagged Action Conditioning} (LAC). 

This layout preserves short sequence length ($L{=}T$) while aligning supervision with causal structure. We predict $a_t$ \emph{given} $i_t$ and $a_{<t}$; since $a_t$ is not in the prefix, there is no self-label leakage. This matches the desired conditional
  $p\!\left(a_t \mid i_t, \mathrm{hist}_{t-1}\right)$ and respects the ``action-given-item'' direction at time $t$. The previous action $a_{t-1}$ enriches the conditioning set for next-item prediction,
  improving $p\!\left(i_{t+1}\mid i_t, a_{t-1}, \ldots\right)$ over item-only input. It is easy to verify that this layout satisfies P1/P2/P3, and in the next section we will discuss with more details why LAC could have better prediction power compared to patched conditioning.

\section{Discussion}
\subsection{Why \textit{LAC} works despite an apparent input mismatch}
The LAC layout feeds \emph{Input} $(i_t,a_{t-1})$ and supervises with \emph{Output} $(i_{t+1},a_t)$. Superficially, this looks misaligned because the action feature in the input is corresponds to the \emph{previous} item. In this subsection, we discuss why such design could be a reasonable approach to fuse action signals at the right causal moment while remaining leakage-safe and compute-efficient.

\subsubsection{Attention head learns the lag-by-one pattern}
Transformers, as very expressive model backbone, could readily internalize a \emph{lag-by-one binding} in a data-driven manner. As $i_{t-1}$ and $a_{t-1}$ co-occur as adjacent tokens with fixed $\text{offset}=1$, gradient descent discovers and amplifies such lag-by-one retrieval/fusion pattern without hand-crafted wiring. At query position $t$, certain head(s) learn keys/queries that place high probability on the source index $t{-}1$ (retrieving a representation carrying $i_{t-1}$ and its context) while the token at index $t$ carries the local channel for $a_{t-1}$. The block then performs \emph{early fusion}:
\[
z_t \;=\; f\!\big(h_{t-1}^{(I)},\, a_{t-1},\, i_t,\, h_{<t-1}\big),
\]
where $h_{t-1}^{(I)}$ is the retrieved item-centric vector and the fusion $f$ is realized by the MLP and short-range attention in transformer layers. This fusion creates a compact joint code of ``what was shown'' and ``how it landed'' before any target-aware head fires, giving both the next-item and action heads access to the same semantically aligned evidence. 

\begin{figure}[h]
 \centering
  \includegraphics[width=0.63\columnwidth]{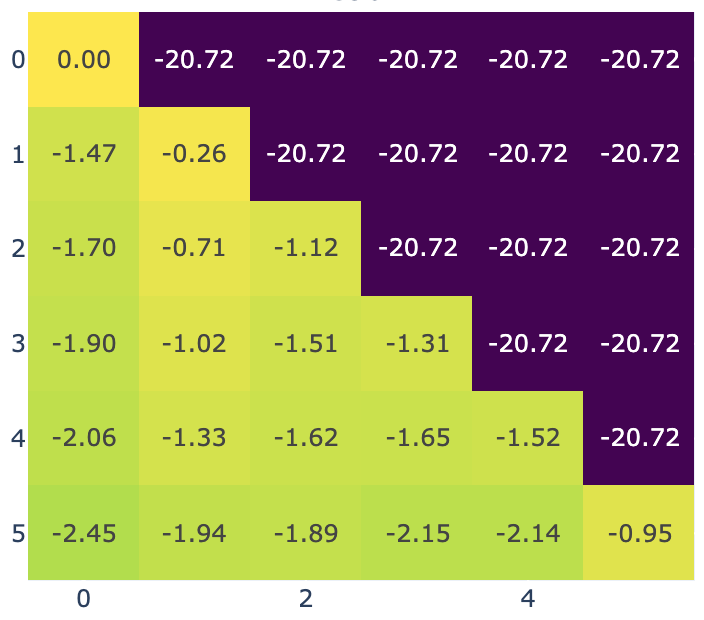}
 \caption{Attention map illustration of the LAC model. We construct a dummy user sequence and visualize last layers attention map. Same with traditional attention map visualizations, X axis presents key index and Y axis presents query index.}
 \label{tab:lag_att_map}
 \end{figure}

To verify that the model does learn about the lag-by-one pattern, we construct a dummy user sequence [$0:unk\_id$, $1:item1$, $2:unk\_id$, $3:unk\_id$, $4:unk\_id$, $5:item1$] and visualize last layer's attention map in Figure \ref{tab:lag_att_map}. One could notice that last token's query pays higher attention to both $1:item1$ and $2:unk\_id$. Though the item token $unk\_id$ at position $2$ is not very informative, position $2$ also has action on $item1$ as input (by design of LAC), since it is right after the $1:item1$ token. This indicates the model attempts to aggregate the information of both $1:item1$ and its associated (lagging) action at position $2$ in the previous layers and fuse the information for final prediction.

\subsubsection{Sufficient statistics and AR(1).}
We also interpret LAC’s effectiveness statistically through two conditionals:
\[
p(a_t \mid i_t,\mathrm{hist}_{t-1}) \quad\text{and}\quad
p(i_{t+1} \mid i_t,a_{t-1},\mathrm{hist}_{t-1}).
\]
The first encodes the business-critical dependency “action given item” at time $t$ without leaking $a_t$ into its own prediction. The second leverages the freshest \emph{pre-treatment} covariate $a_{t-1}$, which summarizes short-term state (binge, fatigue, spending streak) yet is unaffected by the current exposure $i_t$. Information-theoretically,
\[
I\!\big((i_t,a_{t-1});\, i_{t+1}\big)\ \ge\ I\!\big(i_t;\, i_{t+1}\big),
\]
with strict improvement whenever $a_{t-1}$ carries incremental signal about imminent choice. From a time-series perspective, user engagement often exhibits AR(1) structure (most basic model in time series) after modest de-trending
\[
a_t \approx \alpha\,a_{t-1} + g(i_t,\text{state}) + \epsilon_t,
\]
so conditioning on $a_{t-1}$ captures a large fraction of predictable variance. In addition, $a_{t-1}$ is observed at train and test time, yielding short, deterministic gradient paths (attend $t{-}1$ then fuse locally) and reducing the teacher-forcing/exposure bias that arises when training on future, model-predicted variables.

\paragraph{Takeaway.}
What looks like an index mismatch is in fact a causally faithful and optimization-friendly layout of tokens. As a result, early-fused, fully observed evidence serves both retrieval and scoring prediction heads.

\subsection{Why \textit{LAC} could outperform \textit{PC}}

\paragraph{LAC as target-aware attention (early fusion with target item).}
In LAC, at each step-$t$, the model predicts $a_t$ given $i_t$ and the user’s history. Because $i_t$ is part of the step-$t$ token, the attention itself (queries/keys/values in the trunk) is conditioned on the target item. Attention heads can therefore learn query-dependent retrieval patterns such as:
“for \emph{this} item, attend more to prior interactions with similar attributes (e.g., brand/price band/genre).”
This realizes target-aware attention at the representation stage (early fusion), enabling the multi-layer transformer trunk to distill item-specific history into last layer's $h_t$, before any head-specific projection during either training or inference.

\paragraph{PC as target-agnostic attention (late fusion with target item).}
In PC, the trunk produces the last layer's state $h_t$ \emph{without} the target item that conditions action prediction. Only after the stacked layers of transformer does the model inject the item embedding $e(i_{t+1})$ through a lightweight (e.g., MLP) coupler $g(h_t, e(i_{t+1}))$ to estimate $p(a_{t+1}\mid i_{t+1}, \text{history})$.
Thus, target awareness enters \emph{only} in the final prediction head, not in the attention that summarized history. This late fusion is often not expressive enough to reintroduce item-conditioned retrieval patterns, leading to underuse of historical signals for context-dependent actions.

To summarize, the effective layout design of LAC makes target-aware early fusion possible while satisfying the principles P1-P3. Consequently, LAC can yield higher accuracy for action prediction, especially when outcomes rely on subtle, item-specific cues embedded in long user histories.

\subsection{Inference with Parallel Scoring}
In industrial multi-stage recommendation systems (retrieval $\rightarrow$ ranking $\rightarrow$ re-ranking), the ranking stage typically scores $10^3$--$10^4$ candidates per user under tight latency/compute budgets. To meet these constraints while maintaining high MFU (Model FLOPs Utilization) on GPU tensor cores, an effective GR layout should support \emph{one-pass} parallel scoring that reuses the encoded user history instead of recomputing it per candidate. In this subsection, we discuss how both LAC and PC admit one-pass parallel scoring of $C$ candidates during inference, and the difference lies in where target awareness enters and how compute scales with $C$.

\medskip\noindent\textbf{LAC (early fusion via attention).}
Pack a single sequence
\[
[\ \underbrace{\text{history}}_{T}\ \mid\ \underbrace{c_1,\ldots,c_C}_{\text{C}}\ ],
\]
where each candidate token $c_k$ contains that item’s embedding and other contextual features (e.g., request time). Apply a customized block-causal mask:
\begin{itemize}[leftmargin=1.2em]
  \item history tokens use standard causal visibility;
  \item each candidate token attends to all history/itself and no other candidates.
\end{itemize}

We take the hidden state at each candidate token
to score
\[
p(a_{t+1}\mid i_{t+1}{=}c_k,\text{history}\big)\quad\text{for }k=1,\ldots,C.
\]
Per layer, the attention computation is
\[
E(T,C)=\tfrac{1}{2}T^2 + C\,T,
\]
and projection/MLP FLOPs is
$O(T{+}C)$. This yields target-aware parallel scoring with near-linear growth in
$C$ for fixed $T$.

\medskip\noindent\textbf{PC (late fusion at the head).}
PC encodes history once with \emph{target-agnostic attention} to obtain a context $h_t$, then injects each candidate via a lightweight coupler $g(h_t,e(c_k))$ to produce
\[
p(a_{t+1}\mid i_{t+1}{=}c_k,\text{hist}_t) \;=\; \text{Prediction\_head}\big(g(h_t,e(c_k))\big),
\]
for all $c_k$ ($k=1,\ldots,C$) in a single dense, batched GEMM. Complexity is $O(\text{Trunk}) + O(C\,d)$ with the attention cost independent of $C$.

\begin{figure}[h]
 \centering
  \includegraphics[width=0.98\columnwidth]{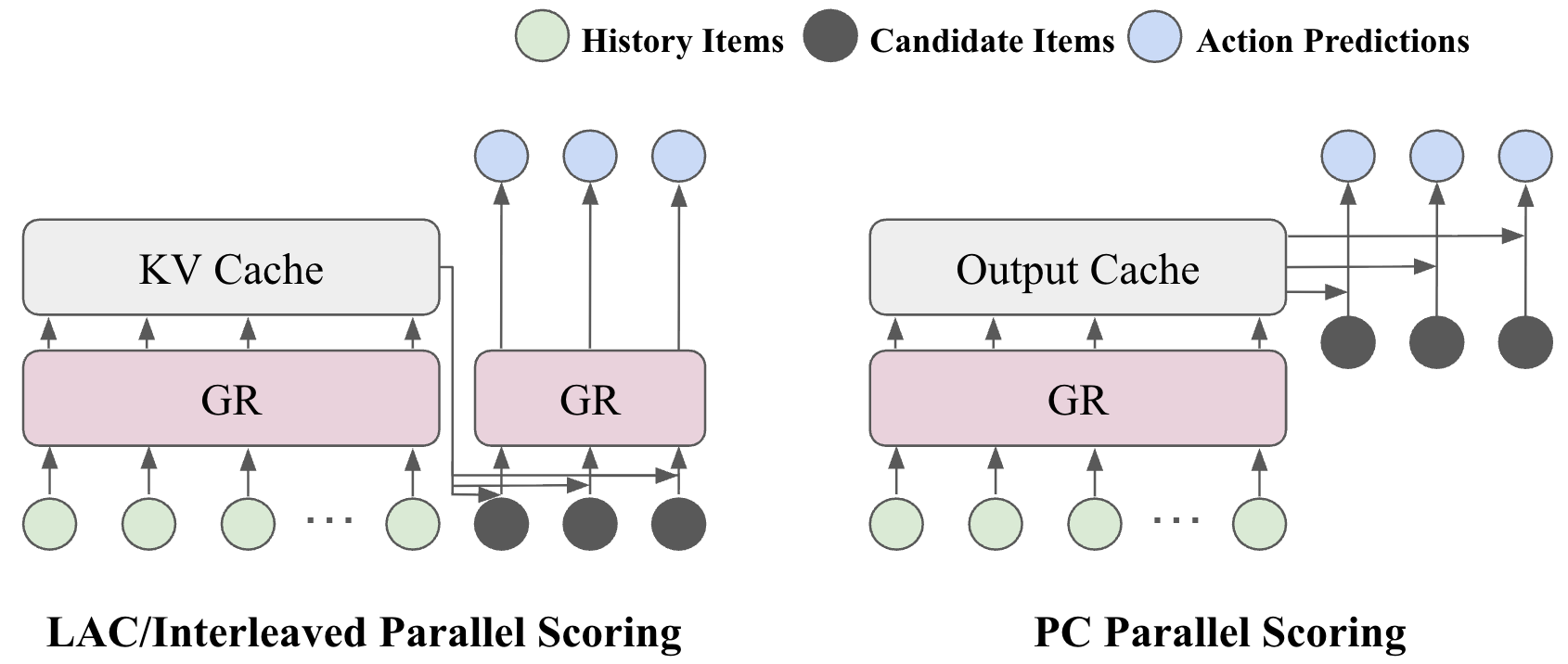}
 \caption{Different inference scoring patterns for LAC and PC. Both approaches allow for parallel scoring of potentially many candidates}
 \label{figure:parallel_scoring}
 \end{figure}

Hence, while both approaches admit one-pass parallel scoring (Figure \ref{figure:parallel_scoring}), LAC provides early, target-aware attention over history with strong action accuracy; PC's late fusion is a light-weight approach at the cost of target awareness appearing only in the head.

\subsection{FLOPs: interleaved vs.\ non-interleaved }
We briefly discuss how much PAC as a non-interleaved approach improves over interleaved in terms of FLOPs, assuming the same autoregressive backbone. Let width $d$, layers $N_\ell$, batch $B$, history length $T$, candidates $C$ (packed scoring with a customized causal mask, no KV-cache). History tokens: $L_h^{\text{non}}=T$, $L_h^{\text{int}}\approx 2T$. Packed length $L=L_h+2C$.

\textbf{Training (full sequence).} Per layer,
\[
\text{FLOPs}_\text{attn}\propto B\,L_h^2\,d,\qquad
\text{FLOPs}_\text{lin/MLP}\propto B\,L_h\,d^2.
\]
Hence
\begin{equation}
\frac{\text{FLOPs}^{\text{int}}_\text{train}}{\text{FLOPs}^{\text{non}}_\text{train}}
\approx \frac{4\,(\text{attn})+2\,(\text{lin})}{1\,(\text{attn})+1\,(\text{lin})}
\in [2,4],
\end{equation}
approaching $4$ when attention dominates (longer contexts), and $2$ when MLP/linear dominates.

\textbf{Inference (parallel scoring, no cache).} Pack a single sequence $[\ \text{history}_{L_h}\ \mid\ i^{(1)},\ldots,i^{(C)}\ ]$
and use a block-causal mask: history tokens are causal; each candidate item $i^{(c)}$ attends to \emph{all} history/itself and to \emph{no} other candidates. The per-layer attention edge count is
\begin{equation}
E(L_h,C)\;=\;\tfrac{1}{2}L_h^2\;+\;C\,L_h,
\label{eq:edges-no-a}
\end{equation}
hence the complexity scales as
\begin{equation}
\text{FLOPs}_\text{attn}\;\propto\;E(L_h,C)\,d,
\qquad
\text{FLOPs}_\text{lin/MLP}\;\propto\;(L_h{+}C)\,d^2.
\label{eq:flops-no-a}
\end{equation}

\noindent\textbf{Interleaved vs.\ non-interleaved.}
Setting $L_h^{\text{non}}=T$ and $L_h^{\text{int}}\approx 2T$, the attention-FLOP ratio is
\begin{equation}
\frac{\text{FLOPs}^{\text{int}}_{\text{attn}}}{\text{FLOPs}^{\text{non}}_{\text{attn}}}
\;=\;
\frac{\tfrac{1}{2}(2T)^2 + C(2T)}{\tfrac{1}{2}T^2 + C T}
\;=\;
\frac{2T^2 + 2CT}{\tfrac{1}{2}T^2 + CT}
\;\in\;[2,4],
\label{eq:ratio-attn-no-a}
\end{equation}
approaching $4$ when history dominates ($T\!\gg\!C$) and $2$ when candidates dominate ($C\!\gg\!T$).
For projection/MLP terms,
\begin{equation}
\frac{\text{FLOPs}^{\text{int}}_{\text{lin}}}{\text{FLOPs}^{\text{non}}_{\text{lin}}}
\;=\;
\frac{2T + C}{T + C}
\;\in\;(1,2],
\label{eq:ratio-mlp-no-a}
\end{equation}
tending to $2$ as $T\!\gg\!C$ and to $1$ as $C\!\gg\!T$.

With packed candidate tokens and parallel scoring, interleaved layouts incur roughly $2$–$4\times$ more attention FLOPs (and up to $2\times$ more linear/MLP FLOPs) than non-interleaved under parallel scoring, with the factor governed by the history/candidate mix.


\section{Experiments}

In this section, we empirically examine how layout choices shape the behavior of generative recommendation models through five research questions:

  \textbf{RQ1}: Compare the proposed \emph{LAC} layout against \emph{Interleaved} (HSTU-style) and \emph{Non-interleaved Patched Conditioning (PC)} across public and industrial datasets, measuring quality and efficiency.

  \textbf{RQ2}: Test the non-interleaved design principles (P1--P3) by evaluating ten carefully constructed layouts and analyzing predicted failure modes.
  
  \textbf{RQ3}: Examine task sensitivity, contrasting tradoffs between multiple action objectives.
  
  \textbf{RQ4}: Study scaling laws under different layouts via sweeps of model size.
  
  \textbf{RQ5}: How do \emph{Interleaved} and \emph{Non-interleaved} layouts differ in \textbf{training} efficiency?

\subsection{Experimental Settings}
\subsubsection{Datasets.} We evaluate using three distinct datasets:
\begin{itemize}
\item  \textbf{Amazon Reviews} \cite{mcauley2015image} \ We use standard Amazon Reviews subsets (e.g., \textit{Beauty}). Following established preprocessing protocols, we apply 5-core filtering, retaining only users and movies with at least five interactions.
\item \textbf{Kuaisar} \cite{sun2023kuaisar} This dataset offers authetic user behavior sequence on a short video platform Kuaishou, which contains more informative user actions including play time. For preprocessing, we also apply 5-core filtering.
\item \textbf{Proprietary Dataset} To validate the effectiveness of the proposed layout on large scale data, we also train and evaluate on our in-house industrial-scale
dataset that consists of tens of millions of users and their behavior sequence on tens of thousands of items, over a period of more than one year. We use the last day of data for evaluation. For scoring task, we select one duration related action and one purchase related action.
\end{itemize}

\begin{table*}[htbp]
\centering
\caption{Performance comparison on public datasets. Boldface denotes the best value while underline indicates the second best result. Metrics marked with \(\downarrow\) are \emph{lower is better}; metrics marked with \(\uparrow\) are \emph{higher is better}.}
\label{tab:public_datasets}
\resizebox{\textwidth}{!}{
\renewcommand{\arraystretch}{1.5} 
\begin{tabular}{l|l|ccccc|ccccc}
\toprule
\multirow{2}{*}{\textbf{Category}} & \multirow{2}{*}{\textbf{Layout}} & \multicolumn{5}{c|}{\textbf{Amazon Beauty}} & \multicolumn{5}{c}{\textbf{Kuaisar}} \\
\cmidrule(lr){3-7} \cmidrule(lr){8-12}
& & \textbf{Rating RMSE \(\downarrow\)} & \textbf{HR@10 \(\uparrow\)} & \textbf{HR@50 \(\uparrow\)} & \textbf{NDCG@10 \(\uparrow\)} & \textbf{NDCG@50 \(\uparrow\)} & \textbf{Play Time RMSE \(\downarrow\)} & \textbf{HR@5 \(\uparrow\)} & \textbf{HR@10 \(\uparrow\)} & \textbf{NDCG@5 \(\uparrow\)} & \textbf{NDCG@10 \(\uparrow\)} \\
\midrule

\multirow{1}{*}{\textit{Interleaved}}
 & HSTU-style & 0.2489 & 0.0861 & \textbf{0.1843} & 0.0497 & \textbf{0.0712} & \underline{0.1020} & \textbf{0.5750} & \underline{0.6770} & 0.4640 & 0.4945 \\
\midrule
 
\multirow{2}{*}{\textit{Non Interleaved}}
 & LAC & \textbf{0.2435} (-2.2\%) & \textbf{0.0869} (+0.9\%) & \underline{0.1816} (-1.5\%)  & \textbf{0.0504} (+1.4\%) & \underline{0.0710} (-0.3\%) & \textbf{0.0990} (-2.9\%) & 0.5621 (-2.2\%) & 0.6534 (-3.5\%) & \textbf{0.4780} (+3.0\%) & \textbf{0.5108} (+3.3\%) \\
 & PC & \underline{0.2480} (-0.4\%) & \underline{0.0863} (+0.2\%) & 0.1807 (-2.0\%) & \underline{0.0499} (+0.4\%) & 0.0703 (-1.3\%) & 0.1020 (+0.0\%) & \underline{0.5698} (-0.9\%) & \textbf{0.6800} (+0.4\%) & \underline{0.4771} (+2.8\%) & \underline{0.5066} (+2.4\%) \\
\bottomrule
\end{tabular}
}
\end{table*}

\subsubsection{Models and Baselines.}
\label{par:models}
For both public and proprietary datasets, we experiment with different layout variants based on our internal generative model, which uses a standard 12-layer GPT-2 style \cite{radford2019language} decoder only transformer with 85M parameters. For the interleaved approach, we follow the layout arrangement from HSTU\citep{Zhai2024ActionsSL}.

\subsubsection{Evaluation Metrics.}
\label{par:metrics}
For the recommendation task on Amazon Beauty and Kuaisar, we adopt the widely used leave-one-out strategy \cite{Kang2018SasRec} to split train, validation and test sets. For our proprietary dataset, we leave out the last day for testing.
We present both ranking task and retrieval task metrics for all datasets. For retrieval related metrics, we use top-k Hit Rate ($HR@k$) and Normalized Discounted Cumulative Gain ($NDCG@k$). We choose different k to accomodate for different dataset characteristics, $k \in \{10, 50\}$ for Amazon and KuaiSAR and $k \in \{1, 5, 300\}$ for proprietary dataset. When calculating retrieval metrics, we default to test against the whole item vocabulary, except for KuaiSAR where we follow \cite{shi2024unisar} and randomly sample 99 negative samples from items user have not interacted with. For ranking tasks, we present Root Mean Squared Error($RMSE$) on chosen action regression tasks. We choose rating for Amazon data and play time for KuaiSAR. For our proprietary dataset we use the above mentiond one duration related action and one purchase related action. For all variants that predict current item's action, we input the current item for action prediction during evaluation.

\subsubsection{Implementation Details.}
\label{par:hyperparams}
For Amazon Beauty and KuaiSAR, we set the maximum input sequence length to 50. We adopt the commonly used sequential recommendation practices in literacture including full shuffle and multi-epoch training. We use a learning rate of 0.001, batch size of 128 and dropout ratio of 0.2. For KuaiSAR, considering the huge vocab size, we used infoNCE loss to replace the original cross entropy loss during training.
For proprietary dataset, the maximum length is set to 1024. We employ a stream training setting, where max tokens per batch is set to 32768. We use a learning rate of 0.0001 and a linear warm-up of 5000 steps.

\subsection{RQ1:Effectiveness}
\textbf{Question.} How does \textbf{LAC} compare to two strong baselines: \textbf{Interleaved} (HSTU-style) and \textbf{Non-interleaved PAC}, across public and internal datasets?

\noindent\textbf{Result.}  We compare three most competitive layout candidates: our proposed LAC approach, PC and industry standard interleaved approach. We show public dataset results in Table \ref{tab:public_datasets} and proprietary dataset results in Table \ref{tab:layout_performance}. For scoring/ranking tasks like action prediction, our LAC approach significantly outperforms other layout variants on both proprietary and public datasets. For retrieval tasks like next item prediction, our LAC approach performs on par with other top candidates.

\RS{1}{LAC approach does empirically outperforms competitive baselines on multiple datasets.}


\subsection{RQ2: Do the non-interleaved principles hold?}
\textbf{Question.} Do P1 (useful context), P2 (action-given-item) and P3 (no leakage) predict performance across a diverse set of \emph{non-interleaved} layouts?

\noindent\textbf{Result.} Comparison between different layout options on the proprietary dataset can be found in Table \ref{tab:layout_performance}. Of all variants, we strictly adhere to principle P3, with no information leakage during training and inference. We can clearly see in the results that adding action information in the input and output phase can both benifit the model performance, which validates our principle P1. We also notice that directly predicting next item's action without item conditioning make the prediction less accurate. Comparing row $9$ with both row $8$ and row $10$, we show that conditioning techniques like PC and LAC can both help with action prediction greatly, which further validates our principle P2.

\RS{2}{By comparing more than ten combinations, we can observe that the proposed principles for non-interleaved layout align with the emprical observations.}

\begin{table*}[htbp]
\centering
\caption{Performance comparison between different layout options on proprietary dataset. Boldface denotes the best value while underline indicates the second best result.}
\label{tab:layout_performance}
\renewcommand{\arraystretch}{1.5} 
\resizebox{\textwidth}{!}{
\begin{tabular}{l|l|ccc|cc}
\hline
\textbf{Category} & \textbf{Layout} & \textbf{HR@1} & \textbf{HR@5} & \textbf{HR@300} & \textbf{Duration Action RMSE} & \textbf{Purchase Action RMSE} \\
\hline

\multirow{1}{*}{\textit{Baseline}}
  & \( I_t \mapsto I_{t+1} \) & 0.2439 & 0.4647 & 0.9186 & - & - \\
\hline

\multirow{2}{*}{\textit{Action only in input}}
  & \( (I_t, A_t) \mapsto I_{t+1} \) & 0.2500 & 0.4694 & 0.9192 & - & - \\
  & \( (I_t, A_{t-1}) \mapsto I_{t+1} \) & 0.2492 & 0.4674 & 0.9188 & - & - \\
\hline

\multirow{3}{*}{\textit{Action only in output}}
  & \( I_t \mapsto A_t \) & - & - & - & 0.2146 & 0.0315 \\
  & \( I_t \mapsto A_{t+1} \) & - & - & - & 0.2423 & 0.0316 \\
  & \( I_t \mapsto (I_{t+1}, A_{t+1}) \) & 0.2444 & 0.4663 & 0.9190 & 0.2454 & 0.0316 \\
\hline

\multirow{4}{*}{\textit{Action in both input and output}}
  & \( (I_t, A_t) \mapsto A_{t+1} \) & - & - & - & 0.2633 & 0.0516 \\
  & \( (I_t, A_{t-1}) \mapsto (I_{t+1}, A_t) \)\,(\textbf{LAC}) & \textbf{0.2516} & \underline{0.4694} & 0.9188 & \textbf{0.1973} & \textbf{0.0307} \\
  & \( (I_t, A_t) \mapsto (I_{t+1}, A_{t+1}) \) & 0.2493 & 0.4690 & \underline{0.9193} & 0.2323 & 0.0311 \\
  & \( (I_t, A_t) \mapsto (I_{t+1}, A_{t+1}) \)\, with \textbf{PC} & 0.2487 & 0.4682 & 0.9190 & 0.2164 & 0.0311 \\
\hline

\multirow{1}{*}{\textit{Interleaved}}
 & HSTU-style & \underline{0.2511} & \textbf{0.4701} & \textbf{0.9198} & \underline{0.1985} & \underline{0.0311} \\
\hline
\end{tabular}
}
\end{table*}


\begin{table}[htbp]
\centering
\caption{Scaling performance of different layout variants. Our proposed LAC still performs on par with more computationally expensive interleaved approach and outperforms PC under a larger scale setting.}
\label{tab:scaling_performance}
\renewcommand{\arraystretch}{1.5} 
\begin{tabular}{l|l|ccc}
\hline
\textbf{Model} & \textbf{Scale} & \textbf{HR@1} & \textbf{HR@5} & \textbf{Duration RMSE} \\
\hline

\multirow{3}{*}{Interleaved}
 & 50M & 0.2511 & 0.4701 & 0.1985  \\
 & 355M & \textbf{0.2639} & \underline{0.4942} & \underline{0.1966} \\
\hline

LAC & 50M & 0.2516 & 0.4694 & 0.1973 \\
(non-interleaved) & 355M & \underline{0.2639} & \textbf{0.4943} & \textbf{0.1945} \\
\hline

PC & 50M & 0.2487 & 0.4682 & 0.2164 \\
(non-interleaved) & 355M & 0.2599 & 0.4922 & 0.2085 \\
\hline
\end{tabular}
\end{table}

\subsection{RQ3: Task sensitivity to layout}
\textbf{Question.} How do layouts behave across multiple scoring tasks (e.g., duration related action vs.\ purchase related action)?

\noindent\textbf{Result.} Since we formulate each tokens prediction task as a multi-task prediction problem, we investigate the efficiency of trade-offs between these targets as a proxy to overall model performance. We fix the relative weight ratio between retrieval(i.e. next item prediction) and scoring/ranking(i.e. action prediction) objectives and tune the relative weight between two scoring tasks. We show the multitask metrics tradeoff of our proposed LAC approach, against the industry standard interleaved approach. Results can be seen in Table \ref{tab:efficient_frontier}. Our proposed LAC approach tend to achieve overall decent tradeoff in multi-task scoring setting, compared to the interleaved approach.

\RS{3}{In multi-task setting, the proposed LAC approach also demonstrates strong trade-offs on potentially conflicting tasks.}

\begin{table}[htbp]
\centering
\caption{Multitask tradeoff of different layout variants.}
\label{tab:efficient_frontier}
\renewcommand{\arraystretch}{1.2}
\setlength{\tabcolsep}{6pt} 
\resizebox{\linewidth}{!}{%
\begin{tabular}{l|l|cc}
\hline
\textbf{Models} & \textbf{W\_duration : W\_purchase} & \textbf{Duration RMSE} & \textbf{Purchase RMSE} \\
\hline
\multirow{3}{*}{Interleaved}
 & W = 1 : 50 & 0.1982 & 0.0304 \\
 & W = 1 : 1   & 0.1978  & 0.0304 \\
 & W = 50 : 1  & 0.1951 &  0.0313 \\
\hline
\multirow{3}{*}{LAC}
& W = 1 : 50 & 0.1979 & 0.0299 \\
& W = 1 : 1 & 0.1967 & 0.0300 \\
 & W = 50 : 1  &  0.1958 & 0.0305 \\
\hline
\end{tabular}%
}
\end{table}


\subsection{RQ4: Scaling under different layouts}
\textbf{Question.} How does layout impact model parameter scaling in GR?

\begin{figure}[h]
 \centering
  \includegraphics[width=0.6\columnwidth]{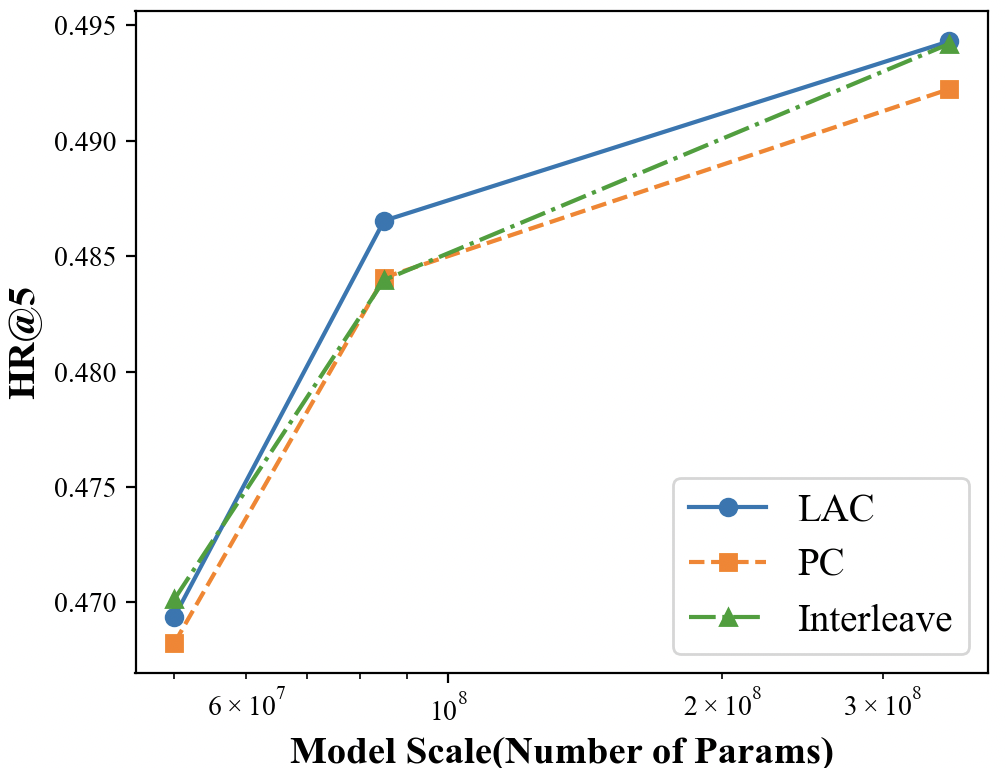}
 \caption{Performance Scaling Curve under different parameter scales. Our proposed LAC outperforms other variants and scales well as we increase model parameters.}
 \label{tab:scaling_curve}
 \end{figure}

\noindent\textbf{Result.} We evaluate the model scaling performance with different parameter sizes ($50M$, $85M$, $355M$), and report the performance scaling curve. As shown in Table \ref{tab:scaling_performance}, our proposed LAC approach scales well with the model complexity and consistently performs on par or even better than the interleaved approach, which is more computationally expensive. Compared with layout candidates of similar computational complexity like PC, our LAC approach performs better at all scales. Visualization of the performnance scaling curve can be found in Figure \ref{tab:scaling_curve}.

\RS{4}{The proposed layout (LAC) consistently has competitive prediction accuracy with different parameter sizes}


\subsection{RQ5: Efficiency of interleaved vs.\ non-interleaved}
\textbf{Question.} How do \emph{Interleaved} and \emph{Non-interleaved} layouts differ in \textbf{training} efficiency under equal hardware and comparable model/data settings?

\noindent\textbf{Result.} We compare the training time the proposed non-interleaved LAC approach and the traditional interleaved approach under different scales. When measuring training cost, we keep all the other parameters the same, including data pipeline settings, max tokens per batch and number of gpus. As shown in Table \ref{tab:training_cost_saving}, compared with the interleaved approach, our LAC approach achieves at least 31.3\% training cost reduction, while achieving similar and even better ranking/retrieval metric performance.

\begin{table}[htbp]
\centering
\caption{Training cost savings of a non-interleaved approach like LAC compared to interleaved baseline at different model scales. We keep all the other training settings the same and only change the layout structure.}
\label{tab:cost_saving_performance}
\renewcommand{\arraystretch}{1.5} 
\begin{tabular}{l|c}
\hline
\textbf{Scale} & \textbf{Training Cost Saving (\%)} \\
\hline
50M & -38.8 \\
355M & -42.7 \\
774M & -31.3 \\
\hline
\end{tabular}
\label{tab:training_cost_saving}
\end{table}

\RS{5}{Non-interleaved approach such as LAC significantly improves the efficiency of generative recommenders.}


\section{Conclusion}

In the paradigm of generative recommendation, interleaved layouts reliably enforce action-given-item conditioning but substantially inflate training and inference costs with lengthened sequences. We recognize action–item layout as an important design axis and develop a methodology for non-interleaved token layout grounded in three principles. Within this framework, we propose Lagged Action Conditioning (LAC), which effectively addresses item–action coupling without doubling length. Experiments on public data and large-scale proprietary data show that LAC achieves superior accuracy–efficiency trade-offs, matching or surpassing interleaved baselines while reducing compute, and outperforming late-fusion alternatives.

\clearpage
\bibliographystyle{ACM-Reference-Format}
\bibliography{reference}

\end{document}